\documentclass[preprint2]{aastex}
\usepackage{spr-astr-addons}
\usepackage{graphicx, epsf, epsfig}
\begin{document}
\title{Bridging the gap between stellar-mass black holes 
and ultraluminous X-ray sources}

\shorttitle{Black hole accretion states}        
\shortauthors{R.~Soria}

\author{Robert Soria \altaffilmark{1}}
\affil{Harvard-Smithsonian Center for Astrophysics, 60 Garden st, 
Cambridge, MA 02138, USA}
\email{rsoria@cfa.harvard.edu}


\altaffiltext{1}{MSSL, University College London, Holmbury St Mary, 
Dorking RH5 6NT, UK}



\begin{abstract}
The X-ray spectral and timing properties of ultraluminous 
X-ray sources (ULXs) have many similarities with the very high state 
of stellar-mass black holes (power-law dominated, at accretion rates 
greater than the Eddington rate). On the other hand, their cool 
disk components, large characteristic inner-disk radii 
and low characteristic timescales have been interpreted 
as evidence of black hole masses $\sim 1000 M_{\odot}$ 
(intermediate-mass black holes).
Here we re-examine the physical interpretation of the cool disk 
model, in the context of accretion states of stellar-mass black holes.
In particular, XTE J1550$-$564 can be considered
the missing link between ULXs and stellar-mass black holes, 
because it exhibits a high-accretion-rate, low-disk-temperature 
state (ultraluminous branch).
On the ultraluminous branch, the accretion rate 
is positively correlated with the disk truncation radius and the 
bolometric disk luminosity, while it is anti-correlated with 
the peak temperature and the frequency 
of quasi-periodic-oscillations. Two prototypical ULXs (NGC\,1313 X-1 and X-2) 
also seem to move along that branch. We use a phenomenological 
model to show how the different range of spectral and timing 
parameters found in the two classes of accreting black holes depends 
on {\it both} their masses and accretion rates. We suggest that 
ULXs are consistent with black hole masses $\sim 50$--$100 M_{\odot}$,
moderately inefficiently accreting at $\approx 20$ times 
Eddington.
\end{abstract}



\section{Introduction}
\label{intro}


Black holes (BHs) are very simple systems, 
entirely characterized by mass and spin. However, 
accretion onto BHs is a much more complex process. 
Observationally, accreting stellar-mass BHs are found in various 
transient states, with different spectral and 
timing properties, and a different balance between 
mass-energy advection, radiation and outflows. 
For BHs in a binary system, accreting from a Roche-lobe-filling 
donor star, the main parameter that determines the accretion 
state is thought to be the mass accretion rate, 
or more precisely the mass inflow rate $\dot{M}$ 
at the outer edge of the accretion disk. This can be larger 
than the accretion rate through the BH horizon, if some 
of the gas is lost in outflows.
The inflow rate depends on the nature 
and orbital separation of the companion star.  
It is convenient to rescale this rate 
as a dimensionless quantity, 
$\dot{m} \equiv \dot{M}/\dot{M}_{\rm Edd}$, 
where $\dot{M}_{\rm Edd} = L_{\rm Edd}/(0.1c^2)$ is the accretion 
rate that would be required to produce a luminosity 
$L = L_{\rm Edd} = 1.3 \times 10^{38}$ erg s$^{-1}$ 
with a standard radiative efficiency $= 0.1$.

BHs accreting at a rate $\dot{m} \la$ a few $10^{-2}$ are generally 
found in the low/hard state (Remillard and McClintock, 2006).
Observationally, this is characterized by a cool, faint thermal 
disk and a hard power-law component (photon index 
$\Gamma \approx 1.5$--$2$). 
The power-law emission comes from the radiatively inefficient 
inner region of the inflow, and is probably due 
to inverse-Compton-scattering in a hot, optically-thin corona 
or at the base of a jet. The disk is sometimes
truncated at a large distance from the innermost stable circular 
orbit. Radio studies suggest (Fender et al.,~2004) 
that most of the accretion power 
is carried out by a steady jet (detected as flat-spectrum 
radio emission), with a possible advected component.

When $0.1 \la \dot{m} \la 1$, accreting BHs are typically 
dominated by an optically-thick, geometrically-thin 
disk, emitting a multicolour blackbody spectrum 
(Shakura and Sunyaev, 1973); this is known as high/soft state, 
or thermal-dominant state (Remillard and McClintock, 2006). 
Most of the accretion power is radiated efficiently. 
Advection is not significant, and jets are quenched.

What happens at $\dot{m} \ga 1$ (very high state, 
or steep-power-law state) 
is less well known. Few Galactic BHs have reached 
this threshold, and only for short periods of time. 
The X-ray spectrum becomes dominated again 
by a non-thermal component, with the disk contributing $\sim 10$--$50\%$ 
of the total flux. As in the low/hard state, the power-law emission 
may be due to Comptonization of the disk photons.
Radiatively driven outflows are also expected, 
and radio flaring is observed in some sources.
It is not clear whether the inner disk is covered 
or replaced by outflows, or by a corona. 
It is also not clear how the accretion power is 
transferred to or directly released in the upscattering 
medium.

A strictly related problem is the interpretation 
of ultraluminous X-ray sources (ULXs). Both their X-ray spectral
and timing behaviour have similarities with the very high state. 
It was also suggested that ULXs may have high accretion rates
$\dot{m} \gg 1$ (Begelman et al., 2006) 
or BH masses much higher than those of Galactic 
stellar-mass BHs (Miller et al., 2004). However, 
a clear understanding of the quantitative connection 
between ULXs and stellar-mass BHs is still missing.
Here, we try to explore this comparison  
and determine whether ULXs are a separate class of BHs, 
or are stellar-mass BHs in a different accretion state.

\section{Thermal disk equations for the high/soft state}
\label{sec:2}

We start by assuming that the disk extends all the way 
to the innermost stable circular orbit, so that 
\begin{equation}
R_{\rm in}  = R_{\rm ISCO} \equiv 6\alpha GM/c^2,
\end{equation}
where $\alpha$ depends on the spin 
of the BH ($\alpha = 1$ for a Schwarzschild BH, $\alpha = 1/6$ 
for an extreme Kerr BH). 
The bolometric disk luminosity is directly related 
to the accretion rate, from general energy-conservation 
principles. Ignoring relativistic corrections,
\begin{equation}
L_{\rm disk} = \frac{GM\dot{M}}{2R_{\rm in}} \approx \frac{1}{12 \alpha} 
     \dot{M} c^2 \equiv \eta \dot{M} c^2, 
\end{equation}
where $\eta$ is the radiative efficiency.

In the standard disk-blackbody model (Makishima et al., 1986, 2000), 
the bolometric disk luminosity is approximated by 
\begin{equation}
L_{\rm disk} = 4 \pi \sigma (T_{\rm in}/\kappa)^4 
     (R_{\rm in}/\xi)^2 \equiv 4 \pi \sigma T_{\rm in}^4 r_{\rm in}^2,
\end{equation}
where: $T_{\rm in}$ is the peak colour temperature; 
$T_{\rm in}/\kappa$ is the peak effective temperature;  
$\kappa$ is the hardening factor ($1.5 \la \kappa \la 2.6$);  
$r_{\rm in}$ is the ``apparent'' inner-disk radius. 
The numerical factor $\xi$ 
was introduced (Kubota et al., 1998) to obtain a correctly 
normalized bolometric disk luminosity, taking into account that 
the fitted peak temperature occurs at $R=(49/36)R_{\rm in}$ 
because of the no-torque boundary conditions.  
With this approximation, and with the choice of $\kappa = 1.7$ 
(Shimura and Takahara, 1995), the physical inner-disk radius $R_{\rm in}$
is related to the apparent value $r_{\rm in}$ by: 
\begin{equation}
R_{\rm in} \equiv (\xi^{1/2} \kappa)^2 r_{\rm in} 
      \approx 1.19 r_{\rm in}. 
\end{equation} 

The bolometric disk luminosity can be obtained from
the observed (or extrapolated) flux:
\begin{equation}
L_{\rm disk} = 2 \pi d^2 f_{\rm bol} (\cos i)^{-1},
\end{equation}
where $i$ is the viewing angle ($i = 0$ means face-on). 
$T_{\rm in}$ can be directly inferred from X-ray spectral 
fitting\footnote{the most commonly used implementation 
of the disk-blackbody model is {\tt diskbb} in {\small XSPEC} 
(Arnaud, 1996).}; $r_{\rm in}$ is derived from $L_{\rm disk}$ 
and $T_{\rm in}$ via Equation (3).

From Eqs. (1), (3), and (4) it follows that:  
\begin{eqnarray}
M  &\approx& \frac{c^2 \xi \kappa^2 \eta}{G(\sigma \pi)^{1/2}} \, 
  L_{\rm disk}^{1/2} \, T_{\rm in}^{-2}\nonumber\\
  &\approx& 10.0 \, \left(\frac{\eta}{0.1}\right)\,
    \left(\frac{\xi \kappa^2}{1.19}\right)\,
    \left(\frac{L_{\rm disk}}{5 \times 10^{38} 
    {\rm{~erg~s}}^{-1}}\right)^{1/2} \nonumber\\
    && \times \left(\frac{kT_{\rm in}}{1  
    {\rm{~keV}}}\right)^{-2} \, M_{\odot},
\end{eqnarray}
which can also be expressed as $L_{\rm disk} \sim M^2 T_{\rm in}^{4}$. 
This is the fundamental evolutionary track of an accretion 
disk in the thermal-dominant state. 
Despite the various approximations, Eq. (6) works well 
(within a factor of $2$) when applied to the masses 
of Galactic BHs, assuming radiative efficiencies $\la 0.2$.

\begin{figure}
\epsfig{figure=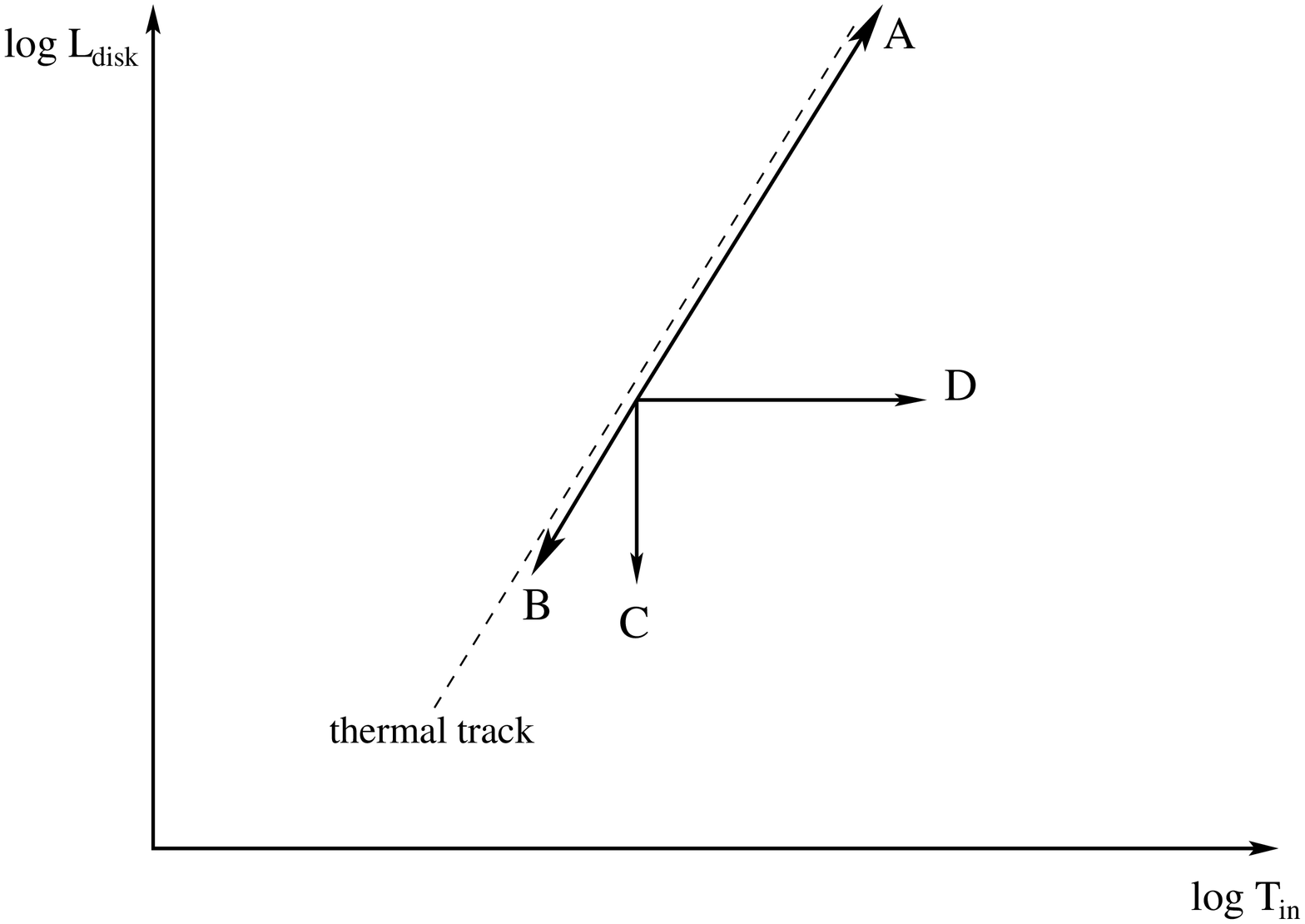, width=8.4cm, angle=0}
\caption{Relation between high/soft state (thermal track 
$L_{\rm disk} \sim T_{\rm in}^{4}$ marked by the dashed line) 
and very high state. Source parameters move 
up along the track (A) when $\dot{m}$ increases; move down 
(B) when $\dot{m}$ decreases or when a fraction of power 
is transferred to an optically-thin corona. 
Partial covering of the disk (C) and hardening of the photon 
spectrum (D) move the source parameters to the right-hand-side 
of the track (very high state). From Soria and Kuncic (2007a).}
\end{figure}

\section{Tracks in the luminosity--temperature plane}
\label{sec:3}



\subsection{Evolution along the thermal track}

As its accretion rate varies, a source in the high/soft
state moves along a track of constant $M$ and $R_{\rm in}$. 
This track typically extends for about one order 
of magnitude in luminosity (e.g., Kubota and Makishima, 2004; 
Miller et al.,~2004). At the upper end, it is truncated 
where the disk luminosity approaches the classical Eddington limit, 
$L_{\rm Edd} \approx 1.3 \times 10^{38} (M/M_{\odot})$ erg s$^{-1}$.
At the lower end, it ends as the source reverts to the low/hard state.

Another process that causes the source parameters to slide along 
the thermal track is the removal of a fraction of power 
$\epsilon > 0$, extracted from the disk 
at each radius (independent of radius). For example, 
this energy may be used to power a corona.
As a result, the peak temperature 
$T_{\rm in} \rightarrow T_{\rm in}' = (1-\epsilon)^{1/4} T_{\rm in}$, and 
$L_{\rm disk} \rightarrow L_{\rm disk}' = (1-\epsilon) L_{\rm disk}$. 
Neither $R_{\rm in}$ nor $r_{\rm in}$ changes;
the emitted spectrum is still a disk blackbody. 
This scenario is equivalent to changing the effective 
accretion rate $\dot{M} \rightarrow \dot{M}' = (1-\epsilon) \dot{M}$.
In summary, changes in the accretion rate, or the (uniform) draining 
of power via non-radiative processes, do not affect the BH mass 
estimate from the fitted disk parameters via in Eq. (6).

\subsection{Evolution off the thermal track}

As the accretion rate approaches or exceeds the Eddington limit, 
an increasing fraction of the photons emitted by the disk 
are mildly upscattered by hot electrons at the disk photosphere. 
This can be modelled by increasing the hardening factor $\kappa$, 
typically from $\approx 1.7$ to $\approx 2.5$ 
(Davis et al., 2005; 2006). 
In the $(T_{\rm in}, L_{\rm disk})$ plane, the effect is to move 
the source horizontally to the right: higher colour 
temperature $T_{\rm in}$ but same effective temperature $T_{\rm in}/\kappa$
and same luminosity $L_{\rm disk}$. The true inner-disk radius 
$R_{\rm in}$ does not change, while the apparent disk radius decreases: 
$r_{\rm in} \rightarrow r_{\rm in}' = [\kappa/(\kappa + \Delta \kappa)]^{2} r_{\rm in}$.
This does not affect the mass estimate, which depends on $T_{\rm in}/\kappa$.

Another effect that needs to be taken into account 
is a partial covering of the disk surface: a fraction $X$ of photons 
may be absorbed or more generally removed from the disk-blackbody 
spectrum by clouds or a moderately optically-thick corona. They 
are re-emitted in other spectral bands or components (e.g., 
X-ray power-law, or infrared). In a self-consistent analysis, 
one must account for all upscattering, downscattering 
and absorption effects, redistributing the (fixed) total available 
accretion power into the various components. But for the purpose 
of relating the disk emission parameters to the BH mass, 
we can simply consider those photons as being lost from 
the disk spectrum. The effect is to reduce the observed disk 
flux and luminosity: 
$L_{\rm disk} \rightarrow L_{\rm disk}' = (1-X) L_{\rm disk}$.
The spectral shape is not altered.
The peak colour temperature $T_{\rm in}$ is not changed. 
$R_{\rm in}$ stays the same but the apparent 
disk radius decreases: 
$r_{\rm in} \rightarrow r_{\rm in}' = (1-X)^{1/2} r_{\rm in}$, 
because $r_{\rm in}$ is indirectly derived from the flux 
normalization. The BH mass inferred from Eq. (6) 
is a factor $(1-X)^{1/2}$ less than the true mass.
 
The combined effect of spectral hardening and Compton upscattering 
is to move the location of the source to the right-hand-side 
of its thermal track, in the $(T_{\rm in}, L_{\rm disk})$ plane (Figure 1). 
Physically, it leads to lower estimates for the fitted radius $r_{\rm in}$ 
(sometimes much lower than the innermost stable orbit), 
and may lead to an underestimate of the BH mass if not properly 
accounted for. As expected, accreting BHs
occupy this region of the parameter space when they are in their
very high state (or steep-power-law state) 
(Remillard and McClintock, 2006), at accretion rates higher 
than in the high/soft state.

\section{The very high state}
\label{sec:4}

\subsection{Very high state in the stellar-mass BH H1743$-$322}

The Galactic BH H1743$-$322 provides a textbook example 
of state transitions (McClintock et al.,~2007). It showed 
a strong outburst between 2003 April and 2003 October 
(Markwardt and Swank, 2003; Miller et al., 2006).
Starting from the low/hard state, it remained hard (power-law dominated)
as it quickly increased in luminosity, until it reached the very high state 
near the peak of its outburst (as monitored by {\it RXTE}). 
It then oscillated between the very high and the high/soft state,
eventually settling in the high/soft state near the high-luminosity 
end of the thermal track. Then, it moved downwards along the thermal 
track (decreasing $\dot{m}$) and finally returned 
to the low/hard state (McClintock et al.,~2007).

Significantly, the disk parameters 
were always on the right-hand-side of the thermal track 
in the $(T_{\rm in}, L_{\rm disk})$ plane (Figures 2 and 3), 
suggesting that we were always seeing a full disk, partly modified 
by Comptonization. There is no indication that the inner boundary 
of the disk retreated to larger radii at any stage, 
either in the low/hard or in the very high state (Figure 4), 
even when most of the X-ray flux was carried out 
by the power-law component.
The disk appeared subluminous for its temperature, compared 
to the thermal track, because some disk photons had been removed 
from the observed thermal component.
Spectral fits give higher peak colour temperatures, 
and lower apparent radii, when the source was power-law dominated. 
This behaviour is typical of other  
(transient) Galactic BHs with low-mass donor stars; 
for example, GRO J1655$-$40 during the 1996 outburst 
(Sobczak et al., 1999).

\subsection{Very high state in ULXs?}

In many respects, ULXs are consistent with being
in the very high state, based on their X-ray spectral 
and timing properties (e.g., Done and Kubota, 2006; 
Dewangan et al., 2006; Goad et al., 2006; 
Strohmayer et al., 2007). 
However, one of the reasons it has been difficult 
to interpret their nature unequivocally is that 
their X-ray spectra can often be fitted equally well, 
in our limited observing window ($\sim 0.3$--$10$ keV), 
with a variety of phenomenological or physical models, 
each corresponding to a very different physical scenario.
Examples of this degeneracy are discussed for example 
in Gon\c{c}alves and Soria (2006), Stobbart et al. (2006), 
Feng and Kaaret (2007). In addition, sometimes the same 
phenomenological fitting model can have two completely 
different physical interpretations.
In the ``standard'' phenomenological model (sometimes known 
as the ``cool disk model''), which seems 
to be applicable to the majority of sources, ULX spectra 
have a dominant power-law component, with a thermal 
(disk?) component contributing $\sim 10$--$50\%$ of the X-ray flux 
(Stobbart et al.~2006). Their fitted temperatures are typically a few times 
lower than in stellar-mass BH disks, and their apparent inner-disk 
radii are two orders of magnitude higher. Spectral hardening 
and partial covering of the disk cannot  
explain this difference. If ULXs lie near or on the right-hand-side
of their thermal tracks, their BH masses ought to be 
two orders of magnitude higher than in Galactic systems. 
This argument has been invoked in support of intermediate-mass BHs 
(Miller et al.,~2004). This {\it physical interpretation 
of the cool disk model} may be unsatisfactory, until independent 
evidence is found for the formation of intermediate-mass BHs
in the local Universe. We need to search for a simpler physical 
interpretation of the cool disk model that does not require 
new astrophysical objects.

\begin{figure}
\epsfig{figure=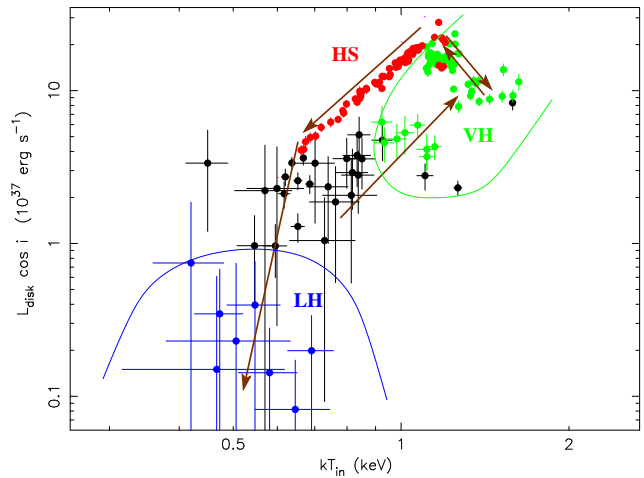, width=8.4cm, angle=0}
\caption{Spectral evolution of the stellar-mass BH 
H1743$-$322 in the temperature--luminosity plane, 
during the 2003 April--October outburst 
(data from McClintock et al.,~2007). The source 
parameters are always located on the right-hand-side of the thermal 
track; this suggests that the disk extends 
to $R_{\rm ISCO}$ even when it is partly modified 
by a corona. Red datapoints mark the thermal-dominant state,  
as identified by McClintock et al.~(2007); green circles 
are for the very high state; blue circles for the low/hard 
state; black circles for intermediate or transition 
states. We assumed a distance of 10 kpc; the viewing 
angle is unknown.}
\end{figure}

\begin{figure}
\epsfig{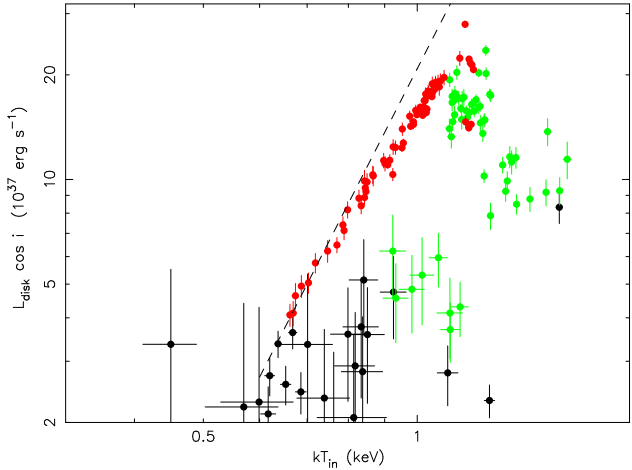}
\caption{Close-up view of the high/soft and very high states 
in the outburst of H1743$-$322; colours are defined as 
in Figure 2. The dashed line marks the thermal track. 
As the accretion rate increases, the source deviates 
more and more from the thermal track (see also Figure 1).}
\end{figure}

\begin{figure}
\epsfig{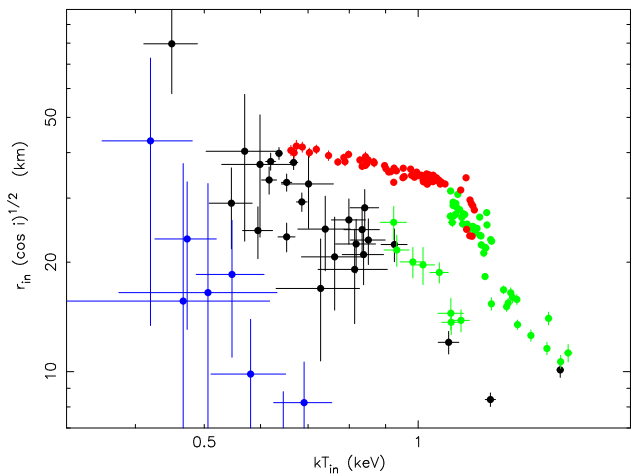}
\caption{Apparent inner-disk radius for H1743$-$322, 
plotted as a function of the peak colour temperature 
of the disk. Colours are defined as 
in Figure 2. In the high/soft state, 
$r_{\rm in} \approx R_{\rm ISCO}$ (Eq. 4). In all other 
states, $r_{\rm in} < R_{\rm ISCO}$, suggesting 
a full disk corrected by a variable hardening factor 
due to Comptonization.}
\end{figure}


\section{Ultraluminous branch in XTE J1550$-$564}
\label{sec:5}

XTE J1550$-$564 is one of the best-known microquasars; 
its BH has a dynamical mass $\approx 10 M_{\odot}$ 
(Orosz et al.,~2002). There is some evidence that the source was 
in a low/hard state at the beginning of the 1998 September -- 1999 
April outburst, just before the initial steep rise 
(Sobczak et al., 2000). In the early phase of the outburst, 
when the integrated X-ray luminosity and presumably the accretion rate 
were at their peaks, the X-ray spectrum was dominated by a power-law, 
with a relatively minor disk contribution. This, and the presence 
of characteristic low-frequency quasi-periodic-oscillations 
(LF-QPOs), are typical signatures 
of the very high state (Remillard and McClintock, 2006).
However, there was an important difference.
During the first (brightest) four weeks of the outburst, 
the source spent some time in the ``normal'' very high state, 
on the right-hand-side of its thermal track, and some time well 
to the left of that track. This is clearly noticeable both 
in the original set of spectral models (Sobczak et al.,~2000, 
which we reproduce here in Figure 5), and in the re-analysis 
of the data done by Kubota and Done (2004) with more complex 
spectral models. When XTE J1550$-$564 was on the left 
of the track, the peak temperature of the disk was much cooler 
than in the normal very high state, and the fitted 
inner-disk radius was much larger than the innermost stable orbit. 
This behaviour has not been seen before 
in other stellar-mass BHs, but may provide a clue 
to understand ULXs. The new state was interpreted (Kubota and Done, 2004; 
Done and Kubota, 2006) 
as a ``strong very high state'' in which the disk may not 
extend to the innermost stable orbit or may be covered 
by a dominant comptonizing corona. We refer to this new state 
as ``ultraluminous branch'' (Figures 5, 6).

At a later stage of the outburst, as the mass accretion rate 
was probably 
declining, XTE J1550$-$564 moved back to the right-hand-side 
of the thermal track, and oscillated a couple of times over the following 
few months between this state and the high/soft state 
(perhaps in response to a moderatley variable but still rather high 
accretion rate). Eventually, the source declined, moving 
downwards along the thermal track. As it re-entered a low/hard state, 
the disk parameters moved again to the left of the thermal track 
(Figure 5, 6). 

\begin{figure}
\epsfig{figure=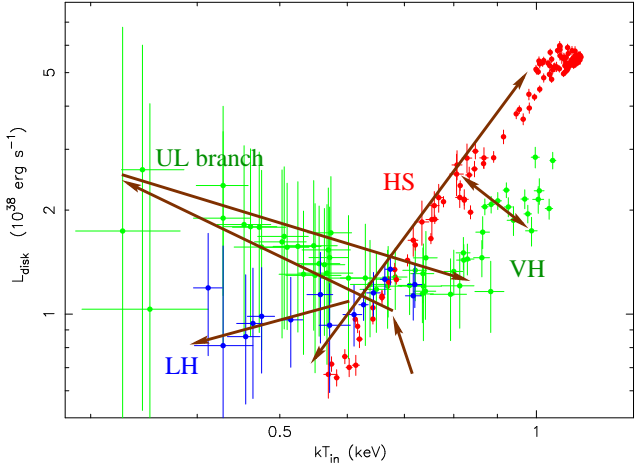, width=8.4cm, angle=0}
\caption{Luminosity-temperature plot for the stellar-mass BH 
XTE J1550$-$564 during its 1998 September -- 1999 April outburst.
Accretion states are colour-coded as in Figure 2. What was 
traditionally classified as the very high state (green 
datapoints) is split here into two states, approximately 
defined as being on the left (ultraluminous branch) 
and on the right (very high state) of the thermal 
track. 
Arrows indicate the approxime evolution track during the outburst. 
The fit parameters are from Sobczak et al.~(2000), and we assumed 
a distance of 5 kpc and a viewing angle $i = 70^{\circ}$.
}
\end{figure}

\begin{figure}
\epsfig{figure=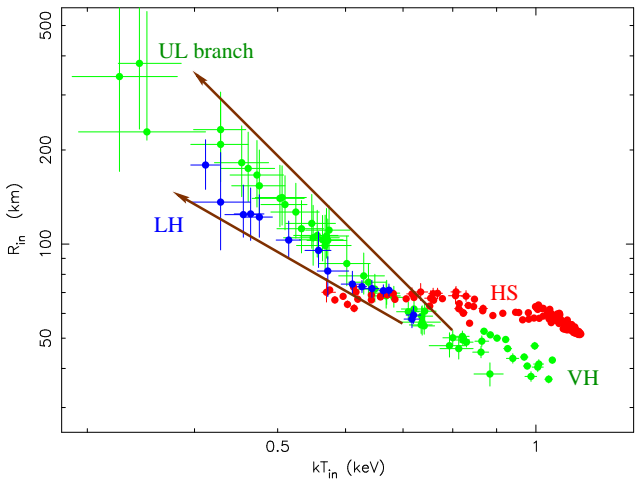, width=8.4cm, angle=0}
\caption{Inner-disk radius of XTE J1550$-$564 
plotted as a function of peak colour temperature. Accretion 
states are colour-coded as in Figure 2. A constant 
conversion factor $R_{\rm in} = 1.19 r_{\rm in}$ 
was used for all datapoints. The radius 
becomes $\gg R_{\rm ISCO} \approx 60$--$70$ km when 
the source is on the ultraluminous branch; 
the disk temperature decreases along the same branch.
An apparently retreating disk is also found in the low/hard state.}
\end{figure}

\section{Physical interpretation of the ultraluminous branch}
\label{sec:6}

Spectral hardening and partial upscattering of the disk photons 
do not explain why the disk should appear larger and cooler, 
at such high X-ray luminosities; one might expect the opposite 
behaviour. Since the BH mass does not change, a large apparent 
radius is generally explained as evidence of a truncated 
or invisible inner disk. Indeed, it has been suggested 
(Done and Kubota, 2006; Dewangan et al., 2006; Goad et al., 2006) 
that when the source is on the ultraluminous branch, 
all the photons emitted from the inner disk are upscattered 
by an optically-thick, low-temperature corona. 
What we still see as direct disk emission 
comes from further out, beyond a transition radius 
$R_{\rm c} \gg R_{\rm ISCO}$, where the disk is also cooler.
Another possibility is that the inner disk is disrupted 
or simply occulted by an optically-thick, radiatively-driven outflow 
(Poutanen et al., 2006; Begelman et al., 2007). 
Alternatively, the inner disk may still be present and directly visible, 
but most of its accretion power is extracted via non-radiative processes, 
such as mechanical outflows or Poynting flux (Kuncic and Bicknell, 2004). 
In this case, the peak temperature occurs much further away 
from the inner disk edge. Some of the power released through  
non-radiative channels would then be converted to power-law photons.
We leave a more detailed discussion of the physical interpretation 
of the transition radius to further work (Soria and Kuncic, 2007b).

An important question to ask is whether the ultraluminous branch 
corresponds to a higher or lower 
accretion state than the classical very high state.
The detected $2$--$20$ keV flux ({\it RXTE} data, Sobczak et al.,~2000) 
was higher in the very high state. However, that is not 
necessarily a good indicator of the total luminosity, 
especially when the disk emission is very soft, nor 
of the accretion rate. For example, the total luminosity 
at the top of the thermal track is $\approx L_{\rm Edd}$, 
with a radiative efficiency $\ga 0.1$. The non-thermal 
processes responsible for the dominant power-law emission 
in the very high and ultraluminous states are expected to have 
lower radiative efficiency. So, the total luminosity 
may still be $\approx L_{\rm Edd}$, just differently 
redistributed between thermal and non-thermal components, 
even if $\dot{m} \ga$ a few. In the case of XTE J1550$-$564, the total disk 
plus power-law flux is at least as high, and probably slightly 
higher, along the ultraluminous branch (Soria and Kuncic, 2007b). 
More importantly, the disk luminosity is slightly increasing 
along the ultraluminous branch, for an increasing inner-disk 
radius. This can happen (recall Eq. 2) only if 
the accretion rates also increases along the ultraluminous branch
(even if the total luminosity did not).
The fact that XTE J1550$-$564 reached both the 
very high state and the ultraluminous branch in the first, 
more energetic phase of its outburst, and only (briefly) 
the very high state in the second, weaker phase 
(Sobczak et al., 2000) also suggests 
that the ultraluminous branch corresponds to a higher 
activity state. As the source started to decline, 
it went from very high state to high/soft state, 
and finally to low/hard state, consistent with 
this interpretation.

\section{Disk parameters on the ultraluminous branch}
\label{sec:7}

A plausible phenomenological way to model 
the disk parameters on the ultraluminous branch is to assume  
a transition radius $R_{\rm c} \gg R_{\rm ISCO}$.
For $R > R_{\rm c}$, the inflow can be approximated 
by a standard disk, such that
\begin{equation}
L_{\rm disk} \approx 4 \pi \sigma T_{\rm c}^4 R_{\rm c}^2,
\end{equation}
where $T_{\rm c} \equiv T(R_{\rm c})$ is the maximum 
(observable) disk temperature.
For $R_{\rm ISCO} < R < R_{\rm c}$, we assume that all the accretion power 
comes out in the power-law-like component, and the standard disk
either is not directly visible, or is disrupted, or emits a negligible 
direct flux.
So, regardless of the details, $R_{\rm c}$ will appear as 
the inner radius when we fit the disk spectrum. 
In this scenario, Eq. (1) no longer holds 
or is no longer relevant to the observed spectrum. It is replaced by 
\begin{equation}
R_{\rm c} = F R_{\rm ISCO}
\end{equation}
with $F = F(\dot{m}) > 1$. 
The visible disk radiates only the accretion power released 
from the outer radius to $R_{\rm c}$. The effective radiative 
efficiency of that part of the inflow 
$\sim M/R_{\rm c} \sim 1/(12\alpha F) < 0.1$. 
Therefore, we have (cf.~Eqs. 2 and 6):
\begin{eqnarray}
L_{\rm disk} &\approx & \frac{GM\dot{M}}{2R_{\rm c}} 
     \approx \frac{1}{12 \alpha F} \dot{M} c^2 
     \equiv \frac{\eta}{F} \dot{m} M c^2 \\
M  &\approx & \frac{10.0}{F} \left(\frac{\eta}{0.1}\right) 
    \left(\frac{\xi \kappa^2}{1.19}\right)\,
    \left(\frac{L_{\rm disk}}{5 \times 10^{38} 
    {\rm{~erg~s}}^{-1}}\right)^{1/2} \nonumber\\
    && \times \left(\frac{kT_{\rm in}}{1  
    {\rm{~keV}}}\right)^{-2} \, M_{\odot} \nonumber\\
   &\approx & \frac{790}{F}  \left(\frac{\eta}{0.1}\right) 
    \left(\frac{\xi \kappa^2}{1.19}\right)\,
    \left(\frac{L_{\rm disk}}{5 \times 10^{39} 
    {\rm{~erg~s}}^{-1}}\right)^{1/2}\nonumber\\
   && \times \left(\frac{kT_{\rm in}}{0.2  
    {\rm{~keV}}}\right)^{-2} \, M_{\odot},
\end{eqnarray}
where the first expression for $M$ is scaled to the typical 
values observed in stellar-mass BHs, and the second one 
is scaled to typical ULX parameters.
An estimate of the BH mass based only on the disk luminosity and
temperature, without taking into account the truncation 
factor $F$, will necessarily over-estimate the BH mass.
We argue that this is the main reason why the intermediate-mass 
BH hypothesis is probably incorrect.

\begin{figure}[t]
\epsfig{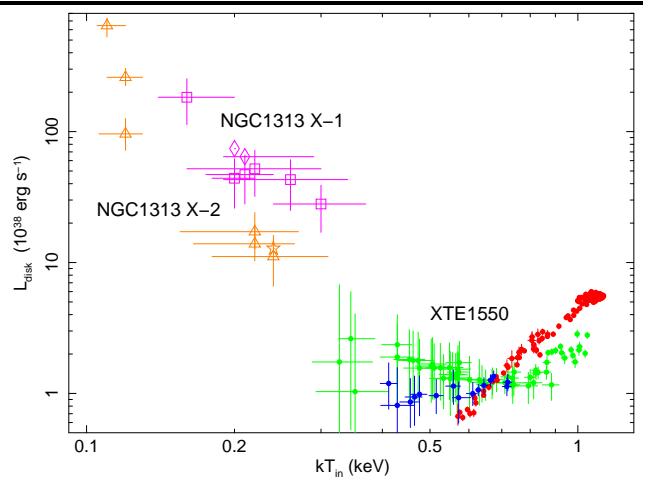}
\caption{As in Figure 5, but with two ULXs (NGC\,1313 X-1 and X-2) 
plotted alongside the stellar-mass BH XTE J1550$-$564.
Both ULXs show an anticorrelation between disk luminosity 
and temperature. This suggests that they are also 
on their respective ultraluminous branches. Unlike 
XTE J1550$-$564, they never occupy any other accretion 
state. The colour coding of the XTE J1550$-$564 datapoints 
is the same as in Figure 2. Datapoints (and, where available, 
error bars) for NGC\,1313 X-1 are plotted in magenta; squares 
are for the {\it XMM-Newton} data (Feng and Kaaret, 2006, 2007), 
and diamonds for the {\it Suzaku} data (Mizuno et al., 2007). 
Datapoints for NGC\,1313 X-2 are plotted in orange (triangles 
for the {\it XMM-Newton} data and a star for the {\it Suzaku} data).}
\end{figure}

Hence, to estimate the BH mass and accretion rate of a source on 
the ultraluminous branch (Eqs. 9 and 10), we need 
to measure  $F= R_{\rm c}/R_{\rm ISCO}$ directly, 
or model its dependence on $\dot{m}$.
For XTE J1550$-$564, we see that the apparent radius 
increases by a factor $F \approx 6$--$8$ along that branch 
(Figure 6). For ULXs, we do not have a complete coverage 
of state transitions, and we do not know the BH mass 
independently. The simplest way to estimate $F$, 
at least as an order of magnitude,  
is to assume that all of the accretion power 
released at $R > R_{\rm c}$ is radiated by the disk, 
and some or most of the power released 
at $R_{\rm ISCO} < R < R_{\rm c}$ is eventually radiated 
as power-law photons, without contributing 
to the thermal disk component. 
From Eqs. (2) and (9), we get
$F \ga L_{\rm tot}/L_{\rm disk}$.  
The inequality sign takes into account the fact 
that the radiative efficiency of the non-thermal processes 
in the inner region is less than 
the efficiency of blackbody emission in a standard disk.
For the most luminous nearby ULXs, 
extrapolating from the relative fluxes of the power-law 
and thermal components in the $0.3$--$10$ keV band, the 
luminosity ratio $L_{\rm tot}/L_{\rm disk} \sim 3$--$20$; in a few other, 
more distant ULXs, there is only a lower limit $\ga 10$ for 
this ratio (Stobbart et al., 2006; Winter et al., 2006).
Hence, we infer that $F \ga 10$ for typical ULXs.

\begin{figure}[t]
\epsfig{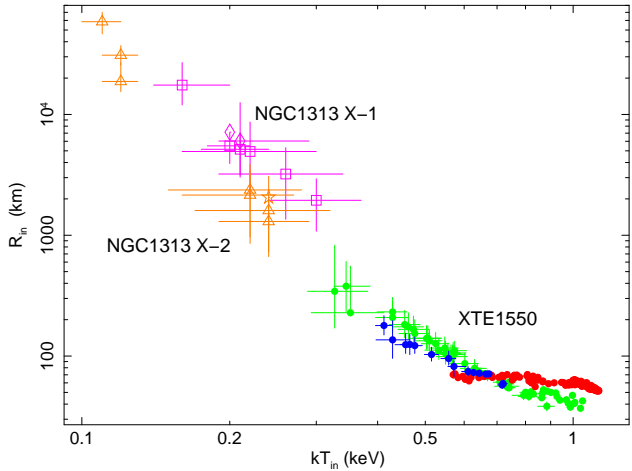}
\caption{As in Figure 6, but with two ULXs (NGC\,1313 X-1 and X-2) 
plotted alongside the stellar-mass BH XTE J1550$-$564. A constant 
conversion factor $R_{\rm in} = 1.19 r_{\rm in}$ was used 
for all datapoints. A viewing angle $i = 60^{\circ}$ was assumed 
for NGC\,1313 X-1 and X-2, for consistency with the flux/luminosity 
conversion of Feng and Kaaret (2006, 2007).
The colour coding is the same as in Figures 7.
The anticorrelation of radius and temperature seen 
in the two ULXs is very similar to (in fact, almost the direct 
extension of) the ultraluminous track of XTE J1550$-$564. }
\end{figure}

\section{Observational predictions}
\label{sec:8}
 
From the fitted disk luminosities and temperatures 
(e.g., Miller et al., 2004; 
Feng and Kaaret, 2005; Stobbart et al.,~2006), 
typical values of $R_{\rm c} \ga 5000$ km are found for many ULXs, i.e., 
$\sim 100$ times larger than typical inner-disk radii 
in stellar-mass BH. If, as we speculate, ULXs are 
accreting BHs on the ultraluminous branch, with $F \ga 10$, 
it means that their BH masses 
are only required to be $\la 10$ times larger 
than typical BH masses in Galactic systems. This simple 
argument suggests that ULX masses are $\la 100 M_{\odot}$. 
At the same time, masses $\sim 50$--$100 M_{\odot}$ 
are still sufficiently high to guarantee that the apparent ULX 
luminosity is comparable or only a factor of a few higher 
than the Eddington limit, and therefore they do not 
require strong collimation of their emission.

The location or evolutionary track of a source in 
the $(T_{\rm c}, L_{\rm disk})$ plane provides 
another observational constraint. To predict the displacement 
of a source from its thermal track when $F>1$, 
we need to understand how $F$, $T_{\rm c}$ and $L_{\rm disk}$
vary as a function of $\dot{m}$. For a fixed BH mass, 
the radiative flux equation of a standard disk tells us 
that $T_{\rm c} \sim R_{\rm c}^{-3/4} \dot{m}^{1/4}$.
It was suggested (Poutanen et al.,~2007; 
Begelman et al.,~2007) 
that $R_{\rm c} \sim \dot{m}$, based on plausible physical 
processes that may form such a transition radius. 
If so, from Eq. (7) we expect that, as the accretion rate 
$\dot{m}$ increases, a source will move along a track 
parameterized by $L_{\rm disk} \approx$ constant, 
$T_{\rm c} \sim \dot{m}^{-1/2}$.
This is not exactly what is found in the {\it RXTE} 
spectra of XTE J1550$-$564, because, there, 
$L_{\rm disk} \sim T_{\rm c}^{-1}$. However, 
the luminosity track is very sensitive to the radial 
temperature distribution 
on the disk, at $R>R_{\rm c}$. Disk models with a distribution flatter 
than $R^{-3/4}$ are sometimes used. For example, it was found 
(Kubota et al.,~2005) that $T \sim R^{-0.7}$ may provide 
a more accurate fit to the X-ray spectral data 
of Galactic BHs. In that case, we expect
$L_{\rm disk} \sim T_{\rm c}^{-0.44}$, with 
$T_{\rm c} \sim \dot{m}^{-0.45}$. For a disk dominated 
by X-ray irradiation (as may be the case if $R \gg R_{\rm ISCO}$), 
$T \sim R^{-0.5}$. In that case, 
$L_{\rm disk} \sim T_{\rm c}^{-4}$, with 
$T_{\rm c} \sim \dot{m}^{-0.25}$. 
Another way to reproduce the observed anticorrelation 
of disk luminosity and peak temperature is to parameterize  
$R_{\rm c} \sim \dot{m}^{\beta}$ with $\beta <1$. For example, 
$L_{\rm disk} \sim T_{\rm c}^{-0.8}$ for 
$R_{\rm c} \sim \dot{m}^{3/4}$ and a standard temperature profile.

Regardless of the precise functional forms of $F(\dot{m})$ 
and $T_{\rm c}(R_{\rm c})$, the significant result is 
that we expect a source to move {\it to the left-hand-side 
of its thermal track}, as the accretion rate increases to 
$\dot{m} \gg 1$. Along that track, the luminosity of the disk component 
stays constant or increases, even if the observed peak 
temperature decreases. 
Thus, it appears that in order of increasing accretion rates, 
the spectral states of an accreting BH are: high/soft (thermal 
track); very high (to the right of that track); ultraluminous 
(to its left). Along this sequence, we also expect the source 
to become more and more dominated 
by non-thermal emission components.



\section{Testing the ultraluminous scenario with fitted ULX parameters}
\label{sec:9}

Let us consider an accreting BH with $M \approx 50 M_{\odot}$.
If $\dot{m} \approx 1$ and $F =1$, 
we expect its spectrum to be dominated by the disk 
component, with a peak temperature $\approx 0.8$ keV, 
a characteristic size of the X-ray emitting region 
$\sim 6GM/c^2 \sim 500$ km, and a disk luminosity 
$\approx 5 \times 10^{39}$ erg s$^{-1}$ $\approx L_{\rm Edd}$. 
This luminosity is consistent with those found in average ULXs, 
but the other fit parameters are not. 
Let us now suppose that $F \approx \dot{m} \approx 20$, 
for the same source on the ultraluminous branch. 
In that case, we expect that the source will still have 
$L_{\rm disk} \approx L_{\rm Edd} \approx 5 \times 10^{39}$ 
erg s$^{-1}$ or slightly higher, but a peak temperature 
$\approx 0.8 \times 20^{-0.5}$ keV $\approx 0.18$ keV, 
a characteristic radius $\sim 20 \times 6GM/c^2 \sim 10^4$ km, 
and an X-ray power-law component with a luminosity 
$> 5 \times 10^{39}$ erg s$^{-1}$. These are more 
typical ULX parameters. If we had observed such a source, 
and had directly inserted the fitted values of $L_{\rm disk}$ 
and $T_{\rm c}$ into Eq. (6) instead of Eq. (10), 
we would have incorrectly interpreted the source 
as an intermediate-mass BH with $M \approx 1000 M_{\odot}$.
The same mistake would have occurred if we had only observed 
XTE J1550$-$564 at the end of its ultraluminous branch. 
In that case, we might have estimated a mass 
$\approx 60$--$80 M_{\odot}$ from Eq. (6), instead 
of $10 M_{\odot}$.

We want to test whether at least some ULXs show evidence 
of spectral evolution along the ultraluminous branch, 
with a predicted (and observed in XTE J1550$-$564) 
anticorrelation between disk luminosity and temperature. 
For this, we consider the two ``classical'' sources, 
NGC\,1313 X-1 and X-2; {\it XMM-Newton} and {\it Suzaku} 
observations of those sources were recently studied 
by Feng and Kaaret (2006, 2007) and Mizuno et al.~(2007).
Their works show that both sources admit at least two alternative 
fitting models: a cool-disk and a hot-disk scenario.
The hot-disk model was considered more physical, partly 
because it would place those ULXs close to the extrapolation 
of a thermal track ($L_{\rm disk} \sim T_{\rm c}^4$), 
while the cool-disk model place them to its left and would imply 
an anticorrelation between disk temperature and 
luminosity\footnote{A caveat about the cool-disk fitting 
model for NGC\,1313 X-1 and X-2 is that the disk temperature 
also appears to be anticorrelated with the column density 
(see in particular Fig.~1 of Feng and Kaaret 2007). 
Thus, it has been suggested that the apparent increase in the disk 
luminosity at lower disk temperatures could be a spurious effect, 
if the anticorrelation between column density and disk temperature 
has no physical basis. However, we argue that an increase 
in the intrinsic absorption as the accretion rates 
increases to more than one order of magnitude above Eddington, along 
the ultraluminous branch, is in fact very plausible. Other 
transient X-ray binaries show higher absorption near their outburst 
peaks. So, we do not think this is a serious reason to consider 
the disk luminosity-temperature anticorrelation unphysical. 
It is, however, possible that the luminosity increase at lower 
temperatures is partly due to uncertainties in the column density, 
and therefore that the slope of the ultraluminous branch is flatter 
than it appears in Figure 7.}. 
However, based on the comparison with XTE J1550$-$564 (Figure 7), 
this is precisely what we expect to find. Therefore, 
we argue that the spectral evolution of those two ULXs 
{\it strongly supports the cool-disk model}. The physical 
interpretation of the cool disk is {\it not} based 
on intermediate-mass BHs, but on a receding 
inner radius $R_{\rm c}$, when the sources are moving 
along their ultraluminous branches at accretion rates 
$\dot{m} \ga 10$.
The increase in the apparent inner-disk radius $R_{\rm c}$ 
is even more obvious when it is plotted against 
$T_{\rm c}$ (Figure 8). We suggest that this is further 
evidence of a fundamental similarity between 
the ultraluminous branch seen in XTE J1550$-$564 
and the ULX behaviour. Note that, unlike XTE J1550$-$564, 
neither NGC\,1313 X-1 nor X-2 have shown evidence 
of ever reaching their thermal tracks (Figures 7, 8), 
which remain undetermined. This makes it more difficult 
to disentangle the effects of higher masses and 
higher accretion rates, and hence to estimate 
the values of $F$ and of their BH masses. 
We will discuss these issues quantitatively and 
in greater details in a forthcoming paper (Soria and Kuncic, 2007b).

\begin{figure}[t]
\epsfig{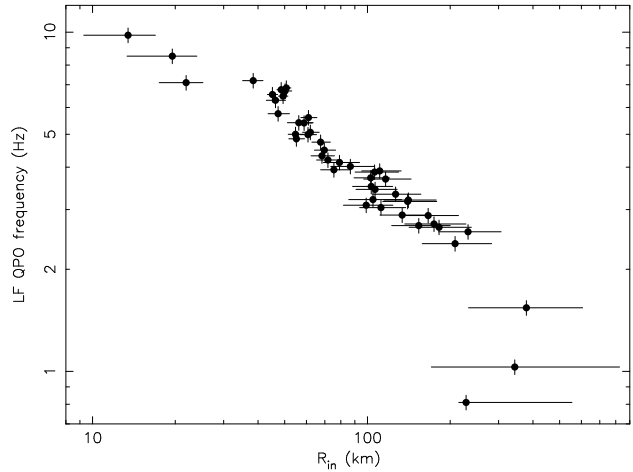}
\caption{Frequencies of the LF-QPOs detected (only) 
in the very high and ultraluminous states 
of XTE J1550$-$564 during its 1998 September -- 1999 April 
outburst, as a function of fitted inner-disk radius.  
A constant conversion factor $R_{\rm in} = 1.19 r_{\rm in}$ 
was used for all datapoints. The anticorrelation of frequency 
and radius (and hence, of frequency and accretion 
rate) provides an independent constraint to ULX masses 
and their accretion rates.}
\end{figure}

\section{Evidence from X-ray timing: low-frequency QPOs}
\label{sec:10}

Some nearby, luminous ULXs show LF-QPOs in their 
power-density-spectra (Strohmayer et al.,~2007; 
Dewangan et al.,~2006; Strohmayer and Mushotzky, 2003).
Such oscillations are more strongly associated with 
the power-law photons (Strohmayer et al.,~2007).
A similar behaviour is observed in stellar-mass BHs, 
including XTE J1550$-$564, in their very high 
and ultraluminous states (Remillard and McClintock, 2006).  
This has been interpreted (e.g., Done and Kubota, 2006) 
as further evidence of a connection between those luminous, 
power-law dominated states in ULXs and stellar-mass BHs, 
also consistent with their spectral properties.
Moreover, it was suggested that LF-QPO frequencies 
scale as $1/M$ across the whole range of BH masses, 
from stellar-mass objects to AGN (McHardy et al., 2006).

LF-QPO frequencies in ULXs are a factor $\approx 20$--$100$ 
lower than in stellar-mass BHs; for example, one of 
the most notable cases is the 20-mHz QPO in NGC\,5488 X-1 
(Strohmayer et al.,~2007). At face value, this appears 
to strengthen the intermediate-mass BH hypothesis 
(Strohmayer et al.,~2007). However, just as we discussed 
for the large inner-disk radii, there may be 
a more physical interpretation, which becomes apparent 
from a comparison with the LF-QPOs in XTE J1550$-$564. 
From the timing analysis of Sobczak et al (2000), 
it is clear (Figure 9) that the LF-QPO frequency in this 
stellar-mass source decreased as the apparent radius increased, 
during the 1998--1999 outburst. That is, 
for a fixed BH mass, the frequency decreased as 
the accretion rate increased along the proposed ultraluminous branch.
The relation is not linear; however, it can be roughly 
approximated as $\nu_{\rm QPO} \sim \dot{m}^{-0.5}$ 
over most of the parameter range. More generally, taking into 
account the expected $(1/M)$ scaling between sources 
of different masses, we can write
\begin{equation}
\nu_{\rm QPO} \sim \dot{m}^{-\delta}\, (1/M)
\end{equation}
with $0.5 \la \delta \la 1$.

This suggests that the lower frequencies found in ULXs 
may also be the result of higher accretion rates, 
not only of higher masses, in agreement with 
our proposed interpretation of the spectral data. 
The different dependence 
of spectral and timing parameters on $\dot{m}$ 
and $M$ suggests that it is in principle possible 
to disentangle the two effects. Preliminary 
back-of-the-envelope calculations suggest that 
accretion rates $\dot{m} \approx 20$ and masses 
$M \sim 50$--$100 M_{\odot}$ are consistent 
with both the spectral and timing data 
of typical ULXs.

In addition to providing an independent constraint 
on the effects of accretion rate and BH mass, 
the observed anticorrelation between 
a disk parameter (inner radius) and the frequency 
of oscillation of the power-law emission component 
is also a fundamental clue to understand the origin 
of such oscillations. One can speculate that LF-QPOs 
originate at the interface of outer disk and inner flow, 
at the transition radius $R_{\rm c}$.
We will present a more detailed analysis of the timing 
results, in the framework of our ultraluminous branch model, 
in Soria and Kuncic (2007b).

\section{Conclusions}
\label{sec:11}

It is often noted that ULXs and stellar-mass BHs 
share X-ray spectral and timing properties 
typical of the luminous, power-law dominated 
accretion state (very high state). 
If we plot the spectral evolution of typical accreting BHs 
in the $(T_{\rm in}, L_{\rm disk})$ plane, the very high state 
is located on the right-hand-side (higher disk temperatures) 
of the thermal track characteristic of the disk-dominated 
high/soft state. 
This would imply that ULXs and stellar-mass BHs 
are entirely separate species, with a two-order-of-magnitude
gap in BH masses between them (intermediate-mass BH scenario).
However, {\it RXTE} spectral fits to the stellar-mass BH 
XTE J1550$-$564 in its 1998--1999 outburst (Sobczak et al.~2000; 
Kubota and Done 2004) 
clearly show that the very high state can in fact be divided into 
two sub-states, located to the right (higher temperature) 
and to the left (lower temperature) of the thermal track, 
respectively. In the latter state (ultraluminous branch), 
the fitted inner-disk radius is much larger than the innermost 
stable circular orbit, and the peak colour temperature much cooler 
than in the high/soft state. This is consistent with a standard 
outer disk truncated or obscured beyond a transition radius 
$R_{\rm c} \gg R_{\rm ISCO}$. The inner region of the inflow 
contributes mostly to the power-law component, perhaps 
through upscattering in a moderately optically-thick corona, 
or in a magnetized wind (Kuncic and Bicknell, 2004). The observed anticorrelation 
between bolometric disk luminosity and peak colour temperature, 
and the observed sequence of spectral state transitions imply 
that the ultraluminous branch occurs at even higher accretion 
rates than the classical very high state, hence 
probably at $\dot{M} >$ a few $M_{\rm Edd}$.

The ultraluminous branch provides {\it a bridge between stellar-mass BHs 
and ULXs}. It supports the cool-disk spectral model for ULXs, 
but not its interpretation in terms of intermediate-mass BHs. 
At the other end of the bridge, we showed that two 
prototypical ULXs (NGC\,1313 X-1 and X-2) also move 
along a track consistent with the ultraluminous branch, 
with anticorrelations between disk luminosity and temperature, 
and between inner-disk radius and temperature.
If this interpretation is correct, all the main spectral and timing 
properties of accreting BHs (including their characteristic 
LF-QPO frequencies) depend simultaneously on two factors: 
a mass scaling and an accretion rate scaling, at least 
when $\dot{M} >$ a few $M_{\rm Edd}$.
We suggest that the observed ULX parameters may be best explained 
with BH masses $\sim 50$--$100 M_{\odot}$ and 
$\dot{M} \approx 20 M_{\rm Edd}$. This also implies 
that the radiative efficiency of the non-thermal medium 
is $\la 0.02$.

In this scenario, ULXs persistently occupy the high-accretion-rate 
ultraluminous branch, whereas stellar-mass BHs would only rarely 
reach such rates, and only near the peak of their outbursts 
(for example, we showed in Section 4 that the stellar-mass BH 
H1743$-$322 never does). We are currently investigating whether brief 
transitions to an ultraluminous branch may have occurred in other 
historical outbursts of Galactic BHs; and whether other 
variable ULXs move along similar tracks.
A significant difference between the two classes of objects 
may be caused by the different types of Roche-lobe 
mass transfer. ULXs may accrete 
a few $M_{\odot}$ over $\la 10^6$ yr from B-type donor stars, 
while transient Galactic BHs accrete from older 
solar-mass stars, which do not persistently 
fill their Roche lobes.

\begin{acknowledgements}
I thank Zdenka Kuncic, Geoff Bicknell, Mark Cropper, Jeroen Homan 
and Jeff McClintock for useful discussions and suggestions. 
In particular, I thank Hua Feng and Phil Kaaret for private communications 
and comments about their study of NGC\,1313 X-1 and X-2, 
and the anonymous referee for various appropriate suggestions.
I also thank Z. Kuncic for funding my visit to Sydney University, where 
part of this work was done. I am grateful to the organizing committee 
of the 5th Stromlo Symposium for their great efforts, that led to a very enjoyable 
and successful meeting. I also thank Jeff McClintock for kindly sharing 
his {\it RXTE} data of H1743$-$322 well ahead of publication. I acknowledge 
financial support from a Marie Curie Outgoing International Fellowship
(No.~509329), from a {\it Chandra}/NASA grant (No.~07620718), 
and from the 2006 RSAA AFL footy tipping competition.
\end{acknowledgements}

\end{document}